\documentclass[submission,copyright,creativecommons]{eptcs}
\providecommand{\event}{M4M 2017} 
\usepackage{breakurl}             
\usepackage{underscore}           

\title{Finite Model Reasoning in Expressive Fragments of First-Order Logic\thanks{This work
		is partially supported by the Polish National Science Centre grant DEC-2013/09/B/ST6/01535.}}

\author{Lidia Tendera
	\institute{Institute of Mathematics and Informatics}
	\institute{Opole University, Poland}
	\email{tendera@math.uni.opole.pl}
}
\def\titlerunning{Finite Model Reasoning in Expressive Fragments of First-Order Logic}
\def\authorrunning{L. Tendera}

\usepackage{amsfonts,url,amssymb,amsmath,color,latexsym,stmaryrd}
\usepackage{tikz}

\newif\ifdraftpaper
\draftpaperfalse

\newcommand{\NPTime}{\mbox{\sc{NPTime}}}%

\renewcommand{\phi}{\varphi} 

\newcommand{\Sat}{\ensuremath{\textit{Sat}}}
\newcommand{\FinSat}{\ensuremath{\textit{FinSat}}}

\newcommand{\ML}{\mbox{ML}}
\newcommand{\MLU}{\mbox{ML}${+}\langle U \rangle$}
\newcommand{\FO}{\mbox{\rm FO}}
\newcommand{\FOt}{\mbox{$\mbox{\rm FO}^2$}}
\newcommand{\FOk}{\mbox{$\mbox{\rm FO}^k$}}
\newcommand{\FOtr}{\mbox{$\mbox{\rm FO}^3$}}

\newcommand{\GF}{\mbox{$\mbox{\rm GF}$}}
\newcommand{\GFt}{\mbox{$\mbox{\rm GF}^2$}}
\newcommand{\GFtEG}{\mbox{$\mbox{\rm GF}^2{+}{\rm EG}$}}
\newcommand{\GFtEQ}{\mbox{$\mbox{\rm GF}^2{+}{\rm EQ}$}}
\newcommand{\FOtEQ}{\mbox{$\mbox{\rm FO}^2{+}{\rm EQ}$}}

\newcommand{\Ct}{\mbox{C$^2$}}

\newcommand{\FL}{\mbox{$\mbox{\rm FL}$}}
\newcommand{\FLk}[1]{\mbox{$\mbox{\rm FL}^{#1}$}}

\newcommand{\UNF}{\mbox{$\mbox{\rm UNF}$}}

\newcommand{\NP}{\textsc{NPTime}}
\newcommand{\PTime}{\textsc{PTime}}
\newcommand{\PSpace}{\textsc{PSpace}}
\newcommand{\ExpTime}{\textsc{ExpTime}}
\newcommand{\ExpSpace}{\textsc{ExpSpace}}
\newcommand{\NExpTime}{\textsc{NExpTime}}
\newcommand{\TwoExpTime}{2\textsc{-ExpTime}}
\newcommand{\TwoNExpTime}{2\textsc{-NExpTime}}

\newcommand{\str}[1]{{\mathfrak{#1}}}

\newcommand{\N}{{\mathbb N}}   

\newcommand{\cutout}[1]{}

\ifdraftpaper
\newcounter{jmc}[section]

\newcommand{\nb}[1]{\textcolor{red}{\bf\large \#}\footnote{\textcolor{blue}{#1}}}
\newcommand{\rev}[1]{\textcolor{magenta}{\bf\large \#}\footnote{\textcolor{red}{#1}}}

\newcommand{\jmf}[1]{\textcolor{jms}\#\footnote{JM: \textcolor{jms}{#1}}}

\newcommand{\ekf}[1]{\textcolor{eks}\#\footnote{EK: \textcolor{eks}{#1}}}

\newcommand{\ltf}[1]{\textcolor{Gray}\#\footnote{LT: \textcolor{lts}{#1}}}

\newcommand{\iphf}[1]{\textcolor{iphs}\#\footnote{IPH: \textcolor{iphs}{#1}}}
\else
\newcommand{\nb}[1]{}
\newcommand{\rev}[1]{}

\newcommand{\jmf}[1]{}

\newcommand{\ekf}[1]{}

\newcommand{\ltf}[1]{}

\newcommand{\iphf}[1]{}
\fi

\newtheorem{theorem}{Theorem}

\newtheorem{lemma}[theorem]{Lemma}

\newtheorem{definition}[theorem]{Definition}

\def\presuper#1#2%
{\mathop{}%
	\mathopen{\vphantom{#2}}^{#1}%
	\kern-\scriptspace%
	#2}

\newcommand{\nc}{\newcommand}
\nc{\bit}{\begin{itemize}}
	\nc{\eit}{\end{itemize}}

\definecolor{Tan}           {cmyk}{0.14,0.42,0.56,0}
\definecolor{Gray}          {cmyk}{0,0,0,0.80}

\begin{document}

\maketitle

\begin{abstract}
Over the past two decades several fragments of first-order logic have been identified and shown to have good computational and algorithmic properties, to a great extent as a result of appropriately describing the image of the standard translation of modal logic to first-order logic. 
This applies most notably to the {\em guarded fragment}, where quantifiers are appropriately relativized by atoms, and the fragment  defined by restricting the number of variables to {\em  two}.
The aim of this talk is to review recent work concerning these fragments and  their popular extensions. When presenting the material special attention is given to decision procedures for 
the {\em finite} satisfiability problems, as many of the fragments discussed contain infinity axioms. We highlight most effective techniques used in this context, their advantages and limitations. We also mention a few open directions of study.

\end{abstract}

\section{Introduction}
Modal logic has good algorithmic and model theoretic properties. 
It is well-known that formulas of propositional modal logics under Kripke semantics can be naturally 
encoded in first-order
logic, using the so-called {\em standard translation}. But since first-order logic is not so well-behaved, in particular the (finite) satisfiability problems are undecidable, it was natural to ask what the right image of the standard translation is and 'Why is modal logic so robustly decidable?' (the last question asked literally by Vardi in \cite{Var97}). 

In order to briefly review some of the answers given, let us have a short look at the standard translation. 
One assigns
to every propositional atom $A$, a unary relation $A(x)$, which is understood as '$A$ is
true in world $x$,' and each reachability relation $R$ corresponds to a binary relation $R(x,y)$. 
This assignment is extended inductively to arbitrary modal formulas: for
every modal formula $\phi$, one inductively defines a first-order formula $tr(\phi,x)$ which
expresses that '$\phi$ is true in world $x$', where the boxes and diamonds are handled by explicit first-order quantification over $R$-accessible points (cf.~\cite{HandbookML}). 
 For example, the modal formula $P \wedge \Diamond ( Q \vee \Box \neg P)$ 
 translates into the following formula
 \begin{equation}\label{eq:eg1}
		Px \wedge \exists y (Rxy \wedge (Qy \vee \forall z (Ryz \rightarrow \neg Pz)))
 \end{equation}

One can observe that the formulas obtained under the standard translation follow some patterns:
(i) variables appear in some fixed order and no rescoping of variables occurs, (ii) quantifiers are relativized by atomic formulas, (iii) negation is applied only to subformulas with a single free variable. These patterns motivated the studies of corresponding fragments of first-order logic defined by appropriately restricting the syntax. 

Moreover, as observed by Gabbay \cite{gabbay1981}, by properly reusing variables one can restrict  their number needed for the standard translation to {\em two}. E.g.~in the previous formula we could replace the variable $z$ by $x$ obtaining:
\begin{equation}\label{eq:eg2}
{Px \wedge \exists y (Rxy \wedge (Qy \vee \forall x (Ryx \rightarrow \neg Px)))}.
\end{equation}
%
%
This  observation is crucial as already the three-variable fragment of first-order is  undecidable,  even for relational signatures featuring only unary nad binary predicates \cite{KMW62}. 

In the next section we introduce the fragments of first-order logic defined by the above mentioned restrictions more formally and we shortly characterize their fundamental properties in terms of the {\em finite} and {\em tree model properties} and in terms of decidability in finite and unrestricted models. Throughout the paper we refer to these languanges as the {\em base languages}. 
In Section \ref{sec:extensions} we review main results concerning satisfiability and finite satisfiability of some popular extensions of the base fragments. In Section \ref{sec:finsat} we sketch a few approaches of proving finite satisfiability for those fragments that do not enjoy the finite model property.

\section{Base languages}

We define the base languages assuming relational signatures not containing any constants or function symbols.
%
%

	\begin{definition}
{\em The two variable fragment}: By the $k$-variable fragment of a logic $\cal L$, denoted ${\cal L}^k$, we mean 
the set of formulas of $\cal L$  featuring at most $k$ distinct variables.
In particular  \FOk{} denotes the set of all first-order formulas with at most $k$ variables. 
 The fragment
 \FOtr{} is already undecidable \cite{KMW62}, therefore, we are
most interested in the two-variable fragment, \FOt.

	\end{definition}

	\begin{definition}{\em The fluted fragment} \cite{purdy:quine76b}:
	Let $\bar{x}_\omega= x_1, x_2, \ldots$ be a fixed sequence of variables.
	%
	%
	%
	We define the sets of formulas $\FL^{[k]}$ (for $k \geq 0$) by structural induction as follows:
	(i) any atom $\alpha(x_\ell, \ldots, x_k)$, where $x_\ell, \dots, x_k$ is a contiguous subsequence of $\bar{x}_\omega$,
	is in $\FL^{[k]}$;
	(ii) $\FL^{[k]}$ is closed under boolean combinations;
	(iii) if $\phi$ is in $\FL^{[k+1]}$, then $\exists x_{k+1} \phi$ and $\forall x_{k+1} \phi$
	are in $\FL^{[k]}$.
	The set of \textit{fluted formulas} is defined as \smash{$\FL = \bigcup_{k\geq 0} \FL^{[k]}$}. A \textit{fluted sentence} is a fluted formula over an empty set of variables, i.e.~an element of $\FL^{[0]}$.
	Thus, when forming Boolean combinations in the fluted fragment, all the combined formulas must have as
	their free variables some suffix of some prefix $x_1, \dots, x_k$ of $\bar{x}_\omega$; and when quantifying, only the last variable in this sequence may be bound. This is illustrated by the fluted sentence in~\eqref{eq:eg1}.  
	\end{definition}

	\begin{definition}{\em The guarded fragment} \cite{ABN98}, \GF{}, 
 is defined as the least set of formulas such that:
(i) every atomic formula belongs to \GF{};
(ii) \GF{} is closed under logical connectives $\neg, \vee,
\wedge, \rightarrow$;
and (iii) quantifiers are appropriately relativised by atoms.
More specifically, in \GF{}, condition (iii) is understood as follows:
if $\phi$ is a formula of \GF{},
$\alpha$ is an atomic formula featuring all the free variables of $\varphi$, and $\bar{x}$ is any sequence of variables in $\alpha$, then the formulas
${\forall} {\bar{x}}(\alpha \rightarrow  \phi)$ and
${\exists}  {\bar{x}}(\alpha \wedge \phi )$ belong to \GF{}. 
In this context, the atom $\alpha$ is called a {\em guard}.
The equality symbol when present in the signature is also allowed in guards.
		\end{definition}

	\begin{definition} 
 {\em The unary negation fragment} \cite{StC13}, \UNF{}, consists of formulas
 in which the use of negation is restricted only to subformulas with at most one free variable. 
More precisely, \UNF{} is defined as  the least set of formulas such that: (i) every atomic formula of the form $R(\bar{x})$ or $x=y$ belongs to \UNF{};  (ii) \UNF{} is closed under logical connectives $\vee$, $\wedge$ and under existential quantification;  
(iii) if $\phi(x)$ is a formula of \UNF{} featuring no free variables besides (possibly) $x$, then $\neg \phi(x)$ belongs to \UNF. 
		\end{definition}

The  base languages are incomparable in terms of expressive power. In particular, the formula  $x\neq y$ is in \FOt{} but not in \UNF. 
Formula (1) lies in the intersection of \FOtr{}  and \FL{}, while (2) is not fluted. Both formulas are guarded and in \UNF{} (the universal quantifier is used as a shortcut in a standard way).
The property: 
\begin{align}\label{eq:eg3}
& \mbox{
	\begin{minipage}{10cm}
		\begin{tabbing}
			No lecturer introduces any professor to every student\\
			$\forall x_1 ($\=$\mbox{lecturer}(x_1) \rightarrow$
			$\neg \exists x_2 ($\=$\mbox{prof}(x_2)
			\wedge$
			$\forall x_3 (\mbox{student}(x_3) \rightarrow \mbox{intro}(x_1,x_2,x_3))))$
		\end{tabbing}
	\end{minipage}
}
\end{align}
belongs to \FL$^3$ but is neither two-variable, nor guarded or in \UNF. 
The property: 
\begin{align}\label{eq:eg4}
& \mbox{
	\begin{minipage}{10cm}
	\begin{tabbing}
	Some node lies on a cycle of length 4\\
	$\exists x_1 \exists x_2( \wedge \exists x_3 (Ex_2x_3 \wedge \exists x_4 (Ex_3x_4 \wedge 
	\exists x_5 (Ex_4x_5 \wedge x_1=x_5))))$
	\end{tabbing}
	\end{minipage}
}
\end{align}
is in \FO$^5$ and in  \UNF, but is neither fluted (the variables in the subformula $x_5=x_1$ do not match the fixed ordering $x_1,\ldots,x_5$) nor guarded (none of the atoms in the subformula $Ex_4x_5 \wedge x_1=x_5$ can be treated as a guard of the quantifier $\exists x_5$).

In the sequel we are concerned with two version of the classical decision problem. For a given logic $\cal L$, 
$\Sat(\cal L)$ is the problem to decide, given a formula $\phi$ of $\cal L$, if $\phi$ is satisfiable. 
Similarly, $\FinSat(\cal L)$ is the problem to decide, given a formula $\phi$ of $\cal L$, if $\phi$ is {\em finitely} satisfiable. i.e.~if it has a finite model. For first-order logic both problems are undecidable \cite{Tur37,Tra50,Tra63} and recursively inseparable \cite{Tra53}. For our base languages the problems are decidable thanks to the {\em finite model property} that we explain below.

\subsection{Finite Model Property and Tree Model Property}

We say that a logic $\cal L$ has the {\em finite model property} (FMP), if every satisfiable formula of $\cal L$ has a finite model.  If $\cal L$ has the finite property then the problems \Sat($\cal L$) and \FinSat($\cal L$) coincide. Moreover, if
$\cal L$ is a subset of first-order logic having the finite model property, then \Sat($\cal L$) (=\FinSat($\cal L$))
is decidable. 

In many cases, the finite model property of some logic comes with a bound on the size of minimal models from which a direct upper bound for the computational complexity of the corresponding satisfiability problem can be derived. (For a given formula it suffices to generate all possible structures within the given size bound and check if any of them satisfies the formula). 

As already mentioned all four of our base languages have the FMP, and hence are decidable. Concerning the bounds on the size of minimal models, \FOt{} has the {\em exponential} model property, and this was the property used in \cite{GKV97} to obtain the tight upper bound on the complexity of the satisfiability problem. 

An algebraic proof of the finite model property for \GF{} can be found in \cite{AndrekaHN99}. In \cite{Gra99} FMP for \GF{} was shown  via the extension property for partial automorphisms of Hrushovski, Herwig and Lascar.  In case of unbounded arities the size of the minimal models that can be easily obtained from this construction is triply exponential in the size of the formula, and not optimal for deciding finite satisfiability.   
FMP for 
\UNF{} was shown  by a reduction to the analogous result for modal logic (which has a very simple proof using filtration~\cite{FischerL79}). 

FMP was used to show decidability of the fluted fragment \cite{Pur96}; the complexity bounds for the bounded variable fragments are not yet tight and they correspond to the best known bounds on the size of minimal models. Namely, the following is known   \cite{P-HST16}:
\bit
\item
\Sat(\FL) is non-elementary; 
\item
\Sat(\FL$^{2k}$) is $k$-\NExpTime-hard and \Sat(\FL$^{k}$) is in $k$-\NExpTime, for all $k\geq 1$. \footnote{There is some very recent work in progress towards showing  that \Sat(\FL$^{2k}$) is $k$-\NExpTime-complete.}
\eit
Tight complexity bounds of the satisfiability problem (= finite satisfiability problem) for the remaining base languages 
are known:
\bit
\item
 \Sat(\FOt) is \NExpTime-complete \cite{GKV97}.
 \item \Sat(\GF)
is \TwoExpTime-complete; 
 \Sat(\GF$^k$) is \ExpTime-complete for
all $k\geq 2$ \cite{Gra99}. 
\item
\Sat(\UNF) is \TwoExpTime-complete; the same holds for  
\Sat(\UNF$^k$) for all $k\geq 3$ \cite{StC13}.
\eit

The optimal complexity bounds for \Sat(\GF) have been obtained by a generalization of the tree model property, known already as an important tool from modal logic. 
  
We say that $\cal L$ has the  {\em (generalised) tree model property}, TMP, iff every satisfiable $\phi\in \cal L$ has a tree (tree-like) model. 
 The fragments \FOt{} and \FL{} do not enjoy the tree model property as they allow to write formulas of the form $\forall x \forall y Rxy$,  enforcing all elements of a model to be connected. 
Gr\"adel showed  \cite{Gra99} that every formula of  \GF{}  with $k$ variables is satisfiable only if it has a model of bounded degree such that the Gaifman graph of this model has tree width at most $k+1$. Similar property holds for \UNF~\cite{StC13}. 

Tree-like models allow the use of powerful tools. For example, in the 
$\mu$-calculus, we can interpret them in the monadic second
order theory of the infinite tree and use Rabin’s theorem (this reduction gives decidability but not good complexity) \cite{Rab69}. The proof of Rabin’s theorem uses tree automata, and by constructing tree automata directly, one 
usually gets good algorithms. However, tree-like models are usually infinite, so TMP is not suitable to decide the finite satisfiability problem. But it might help to improve the complexity bounds, when FMP can be shown independently.


In next sections we will concentrate on logics that do not enjoy FMP and where other techniques to decide finite satisfiability are applied. Before moving on we want to  remark on an important pre-processing phase used in the decision procedures for  the (finite) satisfiability problems. 

\subsection{Normal forms}

When designing algorithms for (finite) satisfiability  in our base languages it is useful to restrict attention to formulas in certain {\em normal forms}. The precise notion depends on the logic but in all cases the normal form formulas  are obtained by iteratively substituting subformulas of the form $\exists y \psi$, for quantifier-free $\psi$ by atoms $R(\bar{y})$, where $R$ is  a fresh predicate letter, $\bar{y}$ denotes the free variables of $\exists y \psi$, and adding appropriate definitions for $R$. Below we recall the corresponding lemmas for \FOt{}, \GF{} and $\FLk{}$.

\begin{lemma}[\cite{GKV97}]\label{lem:FOnf}
For every \FOt{}-sentence $\phi$ one can construct in polynomial time an \FOt{}-sentence $\phi'$ of the form:
$$\phi':=\forall x \forall y \alpha \wedge \bigwedge_{i\in I} \forall x \exists y \beta_i,$$
where $\alpha$ and $\beta_i$ are quantifier-free such that
 $\phi' \models \phi$ and  every model of $\phi$ can be expanded to a model of $\phi'$; moreover, 
if $n$ is the length of $\phi$, then $\phi'$ contains at most $n$ predicate symbols and has length $O(n \log n)$.
\end{lemma}


\begin{lemma}[\cite{Gra99}]\label{lem:GFnf}
For every \GF{}-sentence $\phi$ one can construct in polynomial time a \GF{}-sentence $\phi'$ of the form:
$$\phi':=\bigwedge_{j}  \forall \bar{x} (\alpha_j(\bar{x})\rightarrow \vartheta_j(\bar{x})) 
\wedge \bigwedge_{i} \forall \bar{x} (\beta_i(\bar{x}) \rightarrow \exists \bar{y}  (\beta_i(\bar{y}) \wedge \psi_i(\bar{x},\bar{y})))$$
such that $\phi' \models \phi$ and every model of $\phi$ can be expanded to a model of $\phi'$. Here  the $\alpha_j$, $\beta_i$, $\gamma_i$ are guards and the $\vartheta_j$, $\psi_i$ are quantifier-free; the length of $\phi'$ is linear in the length of $\phi$.
\end{lemma}


%

\begin{lemma}[\cite{P-HST16}]\label{lem:FL-normalform}
	Let $\phi$ be a $\FLk{m}$-sentence over a signature $\sigma$. We can compute, in exponential time, a disjunction $\psi=\bigvee_{k} \psi_k$ over a signature $\sigma'$, where each $\psi_k$ is an $\FLk{m}$-sentence of the form 
$$\psi_k:=\bigwedge_{j} \forall \bar{x} (\alpha_j(\bar{x}) \rightarrow \forall {x'} \beta_j(\bar{x},x'))\wedge
\bigwedge_{i} \forall \bar{x} (\gamma_i(\bar{x}) \rightarrow \exists {x'} \delta_i(\bar{x},x'))$$
such that
$\psi \models \phi$, 
every model of $\phi$ can be expanded to a model of $\psi$; moreover, 
if $n$ is the length of $\phi$, then each $\psi_k$ has length $O(n \log n)$, and $\sigma'$ consists of $\sigma$ together with some additional predicates of arity at most $m-1$. Here, in each conjunct
	$\bar{x}$ is a contiguous seguence $x_1\ldots x_l$ for some $l$ ($1\leq l<m$), $x'=x_{l+1}$, and all of the formulas  $\alpha_j, \gamma_i \in \FLk{[l]}$ and  $\beta_j,\delta_i\in \FLk{[l+1]}$  are quantifier-free.
\end{lemma}

Lemma \ref{lem:FL-normalform} seems weaker than Lemmas \ref{lem:FOnf} and \ref{lem:GFnf} but when aiming at any complexity bound from or above \ExpTime, it also allows one to restrict attention to formulas in normal form (one can consider the disjuncts of $\psi$ one by one and check if any of them is satisfiable).

We also remark that every $\FLk{}$-formula over a signature $\sigma$ consisting of  predicate symbols of arity at most $k$ when transformed to the fluted normal form gives a  formula in $\FLk{k}$. Thus, the fluted formulas obtained by the standard translation from modal logic after  normalization belong to $\FLk{2}$.



\subsection{Historical Remarks} 
The observation that model logic can be seen as a fragment of the two-variable first-order logic was made in 1981 by Gabbay \cite{gabbay1981}. At that time it was known that \FOt{} has the doubly exponential model property and is decidable in \TwoNExpTime{} as shown in 1975 by Mortimer \cite{Mor75}.
The {\em exponential} model property and, hence, tight complexity bounds for \FOt{} were shown by Gr\"adel et.~al.~in 1997 \cite{GKV97}. 
In the same year Gr\"adel, Otto and Rosen published another article \cite{GOR99}, where their performed a test for robust decidability of \FOt{} studying its extensions by adding additional operators corresponding to the operators used in modal logics. This test failed, most of the extensions turned out to lead to undecidable formalism, and therefore \FOt{} was not accepted as the {\em right} image of the standard translation of modal logic (more in the next section). 

One year later  in this context Andr{\'e}ka, van Benthem and N\'{e}meti put forward the guarded fragment \cite{ABN98}. \GF{} does have the the hoped-for nice properties, and has been widely accepted as a {\em better} proposal.   This fragment inspired researchers over the past two decades and brought results having applications in other areas like description logics and database theory. 

\UNF{} is a young fragment, introduced by Segoufin and ten Cate in 2013 \cite{StC13} as an orthogonal (to \GF{}) generalisation of modal logic, that enjoys the same nice properties.  An important additional property of \UNF{} is that it contains unions of conjunctive queries\footnote{A conjunctive query is an existentially quantified conjunction of atoms.}, a class very important in the field of databases. Hence, it is not surprising  that \UNF{} and \GF{} have already been generalised to the {\em guarded negation fragment} that retains the good properties of both \UNF{} and \GF{} logics \cite{BtCS15}.

The origins of the fluted fragment can be traced to a paper given by Quine to the 1968 {\em International Congress of Philosophy}~\cite{purdy:quine69}, in which the author defined what he called the {\em homogeneous $m$-adic formulas}.
In these formulas, all predicates have the same arity $m$, and all atomic formulas have the same argument sequence $x_1, \dots, x_m$. The restriction that all predicates have the same arity is abandoned in~\cite{purdy:quine76b} published in 1976. The history of discovering the decidability and complexity of \FL{} is complicated, and may be a reason why \FL{} has been curiously neglected in the context of our discussion. In particular, 
an earlier claim that \FL{} has the exponential model property \cite{purdy:purdy02} has been just disproved by showing that $\FL$ has the finite model property, and  its satisfiability (= finite satisfiability) problem is decidable, but {\em not} elementary \cite{P-HST16}.

\section{Extensions}\label{sec:extensions}

Modal logic is a very weak formalism in terms of expressive power, however, numerous extensions of ML have been designated to overcome these limitations, leading to extensions that still have good algorithmic properties.
Such  extensions can be defined by  extending the language adding new operators or restricting the classes of frames by adding new axioms. In this section we look at the impact of similar extensions on our base languages. 

\subsection{Additional operators}

We first survey the impact of adding transitive closure operators, (monadic) fixed-points or counting quantifiers that appear, respectively,  in propositional dynamic logic and in temporal logics, in the $\mu$-calculus and in graded modal logics. 
Any of the additional operators implies loss of the FMP that  can be shown be writing {\em infinity axioms}, i.e.~satisfiable formulas that have only infinite models. 

As an example, consider the  $\FOt$-formula with the transitive closure operator, TC, applied to a binary predicate symbol:
\begin{equation}\label{eq:TC-infinite}
\forall x \neg Rxx \wedge \forall x\exists y Rxy \wedge \forall x\forall y (\mbox{TC}(Rxy)\leftrightarrow Rxy).
\end{equation}
This formula is satisfiable and any model of the formula embeds a copy of the natural order relation. 
We can enforce essentially the same property using fixed-points:
\begin{equation}\label{eq:FP-infinite}
\forall x \exists y Rxy \wedge \forall x\forall y \left(Rxy\rightarrow [\mbox{lfp}_{W,x} (Ryx\rightarrow Wy)]x\right). 
\end{equation}
Here the lfp is the set of points that have only finitely many $R$-predecessors. 
A modification of the above examples in the extension of \FOt{} with {\em counting quantifiers,} $\Ct$, can be written as follows:
\begin{equation}\label{eq:fluted-infinity}
\exists x \forall y \neg Ryx \wedge  \forall x \exists y Rxy \wedge \forall x\exists ^{\leq 1} y Ryx.
\end{equation}


Gr\"adel et.~al. studied several  extensions of \FOt{}, in particular the extensions obtained by adding the transitive closure operator and  (restricted) monadic fixed points. In \cite{GOR99}  they showed 
that the extensions of \FOt{} by either transitive closure (in fact, even by transitivity, cf.~next subsection) or fixed points leads to undecidability for both the satisfiability and the finite satisfiability problems. 
Decidability of the satisfiability problem for  \FOt{} with counting quantifiers came as sort of surprise and was shown  independently in 1997 in \cite{GOR97} and  \cite{PST97}. It was also shown that the size of a minimal finite model of a \Ct{}-formula $\phi$ is at least doubly exponential in $|\phi|$ even when the counting quantifiers are only of the form $\exists^{=1}$.  
\NExpTime-completeness of both \Sat{(\Ct)} and \FinSat{(\Ct)} was later proved by Pratt-Hartmann in \cite{PH05}.

The situation with the guarded fragment was different: \GF{} extended with monadic fixed points is decidable and of the same complexity as the base \GF: see Gr\"adel and Walukiewicz \cite{GW99} for the satisfiability problem, and B\'ar\'any and Boja\'nczyk \cite{BaranyB12} for the finite satisfiability problem (note that \cite{GW99} is published in 1999 and \cite{BaranyB12} 13 years later).

Similar properties hold for the unary negation fragment. Despite of the loss of FMP,  decidability and complexity are retained when the fragment is extended by (monadic) fixed point operators \cite{StC13}. To the best of our knowledge, extensions of  \UNF{} by adding transitive closure or counting have not yet been studied.

As for the other two extensions of \GF{}, adding either counting or transitive closure leads to undecidability. 
So special attention has been turned towards the two-variable guarded fragment, where counting quantifiers can be added at no additional cost. Also a decidable extension with restricted transitive closure has been identified in \cite{Michaliszyn09} (finite satisfiability remains open), however the complexity jumps by one exponential in comparison with \GFt{}. These results are summarized in  Table~\ref{tab:extensions}.

\begin{table}
\centering
\begin{tikzpicture}[x=3.2cm,y=1.2cm]
\draw (0,0) grid [step=1] (4,5);
\draw[style=thick] (0,4) -- (4,4);

\node at (1.5,4.5) {Transitive Closure};
\node at (2.5,4.5) {Fixed Points};
\node at (3.5,4.5) {Counting};

\node at (0.5,3.5) {\FOt{}};
\node at (0.5,2.5) {\GFt{}};
\node at (0.5,1.5) {\GF{}};
\node at (0.5,0.5) {\UNF{}};

\node at (1.5,3.5) {undecidable \cite{GOR99}};

\draw[dashed,gray] (1,2.5) -- (2,2.5);
\node at (2.0,3.0) [below left,inner sep=1pt] {\TwoExpTime{} \cite{Michaliszyn09}$^{*)}$};
\node at (1.5,2.0) [above,inner sep=5pt] {\FinSat{}:\quad?};

\node at (1.5,1.5) {undecidable \cite{GOR99}};
\node at (1.5,0.5) {?};

\node at (2.5,3.5) {undecidable \cite{GOR99}};

\node at (2.5,2.7) {\ExpTime{}};
\node at (2.5,2.0) [above,inner sep=2pt] {\FinSat{}:\cite{BaranyB12} \Sat{}:\cite{GW99}};

\node at (2.5,1.7) {\TwoExpTime{}};
\node at (2.5,1.0) [above,inner sep=2pt] {\FinSat{}:\cite{BaranyB12} \Sat{}:\cite{GW99}};

\node at (2.5,0.5) {\TwoExpTime{} \cite{StC13}};

\node at (3.5,3.5) {\NExpTime{} \cite{PH05}};
\node at (3.5,2.5) {\ExpTime{} \cite{PH07}};
\node at (3.5,1.5) {undecidable \cite{Gra99}};
\node at (3.5,0.5) {?};

\draw (0,5) -- (1,4);
\node at (1.0,5.0) [below left,inner sep=5pt] {Extension};
\node at (0.0,4.0) [above right,inner sep=5pt] {Logic};

\end{tikzpicture}
\caption{Overviews of principal extensions of the base languages. The complexity bounds are tight.
	 Key to cells: if not indicated otherwise the values apply to both \Sat{} and \FinSat{} of the corresponding extension. $^{*)}$ only \Sat{} and subject to certain syntactic restrictions.}
\end{table}\label{tab:extensions}


In Table \ref{tab:extensions} we do not list $\FLk{}$ as this kind of extensions have not yet been properly studied. 
In~\cite{purdy:purdy99} the author  considers what he calls {\em extended fluted logic}, in which, in addition to the usual predicate functors, we have equality,  
the ability to exchange arguments in binary atomic formulas and {\em functions} (the requirement that certain specified predicates be interpreted as the graph of a function---a property easily expressed using counting quantifiers). This extension evidently contains infinity axioms, e.g.~formulas equivalent to formula (\ref{eq:fluted-infinity}), hence the claim of \cite{purdy:purdy99} that this extension has FMP is false. 
And it remains open whether $\FLk{}$ with counting, but without the other above mentioned functors, enjoys the finite model property and whether it is decidable.


\subsection{Restricted classes of structures}
In modal correspondence theory various conditions on the
accessibility relations allow one to restrict the class of Kripke structures considered, e.g. to
transitive structures for the modal logic K4, transitive and reflexive---for S4, or equivalence structures for the modal logic S5, and still obtain well-behaved fragments. Also in temporal logics, very natural are classes of structures with some kind of  orderings, where they model time flow. The central condition here is {\em transitivity}. The transitivity axiom is a simple universal first-order formula:
\begin{equation}\label{eq:trans-axiom}
\forall x \forall y \forall z (Rxy\wedge Ryz \rightarrow Rxz)
\end{equation}
however, it is expressible in neither of our base languages, because it contains three variables, has no guard, and the atom $Rxz$ is not fluted. Moreover, adding transitivity axioms allows one to write sentences that have only infinite models (e.g.~replacing the last conjunct in the formula (\ref{eq:TC-infinite}) by the transitivity axiom (\ref{eq:trans-axiom})). 

Hence, the question therefore arises as to whether transitivity (or related properties like orderings or equivalence relations) could be added at reasonable computational cost. We have already seen that it cannot be done in general. 
In the past years 
various extensions of \FOt{} and \GFt{} were investigated
in which certain distinguished binary relation symbols are declared to denote transitive relations, equivalence relations, or linear orderings. It turns out that the decidability of these fragments usually depends on the {\em number} of the distinguished relation symbols available. 

%
  For three linear orders,
 both satisfiability and finite satisfiability are
 undecidable~\cite{Kie2011,Otto01}. Similarly for three equivalence relations~\cite{Kie05}. Turning to transitive relations,
 the satisfiability problem becomes undecidable for both 
 satisfiability and finite satisfiability of \FOt{} in
 the presence of two transitive relations (or even in the presence of one transitive relation and one equivalence relation \cite{KT09}). 

 The complexity bounds for such decidable 
 extensions of \FOt{} and \GFt{} are in many cases identical, a notable
 exception being the case of two equivalences, which,
 for \GFt{} yields a \TwoExpTime-complete logic~\cite{Kie05}, and for
 \FOt{}---a \TwoNExpTime-complete logic~\cite{KMP-HT14}.  Table \ref{tabela}
 summarizes the above results. We do not list there extensions of \GFt{} with linear orders, as linear orders actually destroy the guardedness of a logic: any pair of elements is guarded by a
 linear order, and the results from \FOt{} with linear orders can be applied to \GFt.

\begin{table}[th]
	\begin{center}
		
		\small\hspace{0mm}  
		\begin{tabular}{|c||c|c|c|c|}\hline
			& & \multicolumn{3}{|c|}{} \\
			{\large{Logic}  }  & {{Special symbols}} &
			\multicolumn{3}{c|}{{Number of special symbols in the signature}} \\
			\cline{3-5}
			&  &  1 & 2 & 3 or more \\
			%
			%
			\hline
			\hline &  & & & \\
			\Large{\bf \GFt{}}  & Transitivity  & \TwoExpTime   & undecidable           & undecidable  \\
			&                       & \Sat{}: \cite{Kie05} \FinSat{}: \cite{KT07,KT-FinSatGFTG}  & \cite{Kie05,Kaz06}    & \cite{GMV99} \\
			
			%
			\cline{2-5}
			%
			%
			\cline{2-5}
			FMP  & & & & \\
			\ExpTime & Equivalence  &FMP, {\NExpTime} & {\TwoExpTime} & undecidable\\
			\cite{Gra99} &                       &\cite{KO05}& \cite{KP-HT15}  & \cite{KO05}\\
			\hline\hline
			& & & &\\
			{\Large{{ \FOt{}}}} & Transitivity & {{{ in \TwoNExpTime}}} \cite{ST13}$^{*)}$  & undecidable & undecidable \\
			&              & \FinSat{}: ?           & \cite{Kie05,Kaz06} & \cite{GOR99} \\
			\cline{2-5}
			FMP \cite{Mor75} & & &   & \\
		
			\NExpTime      &  Linear order &  \NExpTime    & \Sat{}: ?  &  undecidable\\
			\cite{GKV97}    &               & \cite{Otto01} & {\sc ExpSpace}$^{**)}$ \cite{SZ12} & \cite{Otto01,Kie2011} \\
			\cline{2-5}

			& & & & \\
			& Equivalence  & {FMP}, \NExpTime & {\TwoNExpTime}    & undecidable \\
			&              &  {\cite{KO05}}   & \cite{KMP-HT14}       & \cite{KO05}\\
			\cline{2-5}
			
			\hline
			
		\end{tabular}
	\end{center}
	\caption{{Overview of two variable logics over restricted classes of
			structures. Unless
			indicated otherwise, the complexity bounds are tight. Key to
			symbols: $^{*)}$ only general satisfiability and for a restricted variant
			$^{**)}$ only finite satisfiability and subject to
			certain restrictions on signatures.}}\label{tabela}
\end{table}

For
\GFt{}, it also makes sense to study variants in which the
distinguished predicates may appear only in guards \cite{GMV99}. In
this case, \GFt{} with {\em any} number of equivalences appearing only
as guards remains \NExpTime-complete \cite{Kie05}, while \GFt{} with
{\em any} number of transitive relations appearing only as guards is
\TwoExpTime-complete \cite{ST01,Kie03} (tight complexity bounds for the finite satisfiability problem are established in \cite{KT-FinSatGFTG}).   


The properties of \UNF{} and $\FLk{}$ over restricted classes of structures have not yet been investigated. Obviously, decidability results for extensions of $\FOt{}$ imply decidability of the same extensions of $\FLk{2}$ or \UNF$^2$. 
Also it is not difficult  to see that the undecidability result for \FOt{} with three equivalence relations can be adapted to the fluted case, hence the satisfiability and the finite satisfiability problems for  $\FLk{2}$ with at least three equivalence relations is undecidable. Other cases need more detailed inspection, additional research, perhaps also novel techniques. 


It is clear that classes of structures defined by stipulating that some binary predicates satisfy some universal first-order formula by no means exhausts the relevant possibilities. One may as well consider e.g.~well-founded structures, trees or forests;  notions not expressible in first-order logic which arise naturally in a wide range of
contexts. 
Also the world of guarded logics is much richer than shown above. E.g.~more liberal guardedness conditions (loosely- or clique-guarded, packed fragment) and guarded fragments of other logics have been studied (guarded second order logic, Datalog LITE). 

\section{Deciding FinSat}\label{sec:finsat}

Before we review some techniques used for solving the finite satisfiability problem for logics without FMP we first notice a few potential difficulties. 

Let $\phi$ be the following formula, where $P_0, \ldots, P_{n-1}$ are unary predicates and $R$ is a binary predicate:
\begin{equation}\label{eq:trans-finite}
\phi= \exists x P_0x \wedge \bigwedge_{0\leq i <n} \forall x (P_ix \rightarrow \exists y (Rxy \wedge P_{i+1}y))\wedge 
\bigwedge_{0\leq i<j<n} \forall x \neg (P_ix \wedge P_jx).
\end{equation}
The formula $\phi$ has a simple infinite model that is an $R$-chain of elements on which the unary predicates alternate. In order to get a finite model the  $R$-chain must close into cycles. By using combinations of the unary predicates to encode a binary number at a given point of a model, one can easily enforce those cycles to have exponential length w.r.t.~the length of the formula. If additionally $R$ is declared transitive, these cycles induce $R$-cliques. Note that $\phi$ is  a formula in all our base languages. 

Smallest finite models might also be relatively large w.r.t.~the length of the formula used to define them (and also in comparison to the optimal upper complexity of the algorithms deciding  finite satisfiability). Recall e.g.~the example from \cite{GOR97}, where a family of finitely satisfiable  \Ct{}-formulas
$\{\phi_n\}_{n \in \N}$ over a signature with one binary  and $n$ unary predicate symbols is given, such that
every finite model of $\phi_n$ contains an isomorphic copy of a full binary tree of height $2^n$  and $\phi_n$ has length $O(n\log n)$. Hence every model of $\phi_n$ has size at least $2^{2^n}$.  We remark at this point that both \Sat{(\Ct{})} and \FinSat{(\Ct{})} are \NExpTime-complete.

This suggest that when designing efficient algorithms for the finite satisfiability problem one can not rely on properties of unrestricted models or on direct constructions of models of minimal size.

In this context let us also mention two titles of papers from the DL community praising unrestricted reasoning versus finite reasoning: 'Nominals, inverses, counting, and conjunctive queries or: Why infinity is your friend!' \cite{RudolphG10} and 
'The curse of finiteness: Undecidability of database-inspired reasoning problems in very expressive description logics' \cite{Rudolph16}.

\subsection{More or less natural reductions}

A perhaps most natural approach to establish (un)decidability or tight complexity bounds of some logic is to reduce formulas of one logic to another one. This classical approach can be illustrated by the extension of \UNF{} by fixed points, UNFP. In fact in \cite{StC13} an exponential reduction from UNFP to the modal $\mu$-calculus is presented that additionally preserves finiteness of the models. This immediately gives $\TwoExpTime$-upper bounds for the complexity of both the satisfiability and the finite satisfiability problems. 
The same reduction allows one to deduce also FMP of \UNF{} (under this reduction a formula from \UNF{} translates to a modal formula without fixed points) and TMP of UNFP.

Another natural idea of solving the finite satisfiability problem for a logic that has a decidable satisfiability problem, might be to reduce the first problem to the later. This concept has an additional advantage, as for unrestricted
reasoning and a family of some simple logics there is a wide range of applicable algorithms (such as e.g.~tableau algorithms that rely on TMP or resolution calculi), which often perform well in practical implementations.

This approach has been investigated by Rosati in \cite{Rosati08} for a relatively inexpressive logic called 
DL-Lite$_{\cal F}$ that already lacks FMP. The idea has been later extended by Garcia~et.al.~to the logic Horn-{\cal{ALCQI}} \cite{GarciaLS14}. 
It requires additional research to find out if this concept can be further extended to non-Horn logics. 

\subsection{Finitary unravellings and locally acyclic structures}

When we are concerned with a decidable logic that does not have FMP but has TMP, a natural idea is to study finitary unravellings, obtained by 'bending' some edges in the tree-like models to keep the structure finite but at the same time similar to a tree, i.e.~acyclic. Here, when saying that a structure is {\em acyclic} we mean that its {\em hypergraph} is. 
We have already observed that it is not always possible for a given formula $\phi$ to get a model of $\phi$ that is at the same time finite and acyclic, cf.~the formula in  (\ref{eq:trans-finite}). 

To address the above idea a notion of {\em $k$-acyclic} structures, where $k$ is some parameter, is introduced. Informally, in a $k$-acyclic structure $\str{A}$ there are no cycles of length at most $k$; more precisely, every induced sub-hypergraph of the hypergraph of $\str{A}$ of up to $k$ vertices is acyclic. 

The aim then is roughly to show that if a formula $\phi$ has a finite model then $\phi$ has a $k$-acyclic model, where $k$ depends only on $\phi$. Having such a property in hand, one can restrict attention to locally acyclic structures, that are usually easier to handle. This approach has been introduced by Otto \cite{Otto04} for \GF{} over restricted signatures and later extended in \cite{BGO13} to full \GF{}, showing that every finite structure is \GF-bisimilar to a finite structure whose hypergrah is locally acyclic. As an application of the general result a new proof of the (small) finite model property for \GF{}  with optimal bounds on the size of minimal models is obtained.

We remark that the above results underlie the correctness of the reduction outlined in the previous subsection from UNFP to the $\mu$-calculus in the finite case. They are also one of the main ingredients of the decidability proof for the finite satisfiability problem for the extension of \GF{} with fixed points \cite{BaranyB12}. 


\subsection{Deciding (Fin)Sat by reduction to linear or integer programming}


Here we briefly describe a less direct approach that has successfully been applied to extensions of \FOt{} to get optimal complexity bounds when the logic allows one to formulate sentences that have relatively large finite models w.r.t.~optimal complexity bounds for (finite) satisfiability. 

The brief  idea is to identify (finitely many types of) building blocks of a potential model and connecting conditions for them, and describe them in a {\em succinct way}. It turns out that these conditions often can be described by a set of (in)equalities. 
In such cases the approach has an additional advantage, namely it allows one to solve simultaneously both $\Sat({\cal L})$ and $\FinSat({\cal L})$: in case of $\FinSat({\cal L})$ given $\phi\in {\cal L}$ we look for solutions of the corresponding equation system over $\N$, in case of $\Sat({\cal L})$ we look for solutions over so-called extended integers, $\N \cup \{\aleph_0\}$.\footnote{E.g.~the equation $x+1=x$ has no integer solution, but has a solution over extended integers $x=\aleph_0$. If such equation appears positively in the conditions describing models of a formula $\phi$ we  deduce that $\phi$ has no finite models.} Moreover, this approach does not depend on TMP and gives hope to think about practical implementation using existing linear/integer programming solvers. 

This approach has been applied to establish optimal upper complexity bounds for an expressive description logic with (restricted) counting quantifiers in \cite{LST05} (\ExpTime), for \Ct{} in \cite{PH05} (\NExpTime), and for the quarded fragment of \Ct{} in \cite{PH07}  (\ExpTime). The (N)\ExpTime-upper bounds should be contrasted with the remark that in these logics the size of minimal models is doubly exponential in the size of the formula.



The linear/integer programming approach has been made more transparent in \cite{KMP-HT14}
when dealing with the extension of \FOt{} with two equivalence relations,  \FOt+$\{E_1,E_2\}$. 
Suppose  $E_1$ and $E_2$ are the equivalence symbols in the signature. The strategy employed in \cite{KMP-HT14} starts with the observation that the {\em intersections} (i.e.~equivalence classes of the coarsest common refinement $E_1 \cap E_2$ of the equivalence relations) arising in any
model of a formula $\phi$ could, without loss of generality, be assumed to have cardinality exponentially bounded as a function of the size of $\phi$. In any such model, every $E_1$-class, and also every $E_2$-class, is the union of some set of such 'small intersections'; and any given $E_1$-class and $E_2$-class are either disjoint, or have exactly one common intersection. This decomposition into equivalence classes allowed one to picture such a model as an edge-coloured, bipartite graph: the $E_1$-classes are the
left-hand vertices; the $E_2$-classes are the
right-hand vertices; and two vertices are joined by an edge just in case they share an intersection,
with the colour of that edge being the isomorphism type of the intersection concerned.

Evidently, the formula $\phi$ imposes constraints on the types of intersections that may arise, and on how intersections
may be organized into $E_1$- and $E_2$-classes; and it was showed in~\cite{KMP-HT14} how
these constraints translated to conditions on the induced bipartite graph of equivalence classes. 

In this way, the original (finite) satisfiability 
problem for \FOt+$\{E_1,E_2\}$ was nondeterministically reduced to the problem of determining the existence of a (finite)  edge-coloured bipartite graph satisfying certain conditions on the local configurations it realizes. The latter problem was called BGESC (for 'bipartite graph existence with skew constraints and ceilings').  By showing BGESC and its finite version to be \NPTime-complete, an optimal \TwoNExpTime-upper bound for both the satisfiability and the finite satisfiability problems for \FOt+$\{E_1,E_2\}$ was obtained. Membership in $\NPTime$ for both BGESC and the finite BGESC problems was shown by a nondeterministic polynomial reduction to an integer-programming problem. 

In \cite{KMP-HT14} and later in \cite{KP-HT15} two simpler variants of the BGESC problem were introduced called, respectively, BGE and BGE$^*$. They were shown to remain in $\PTime$ via polynomial  reductions to the linear programming problems (for the finite versions) and reductions to the satisfiability problem for propositional Horn clauses (for the unrestricted versions). Reductions to the (finite) BGE$^*$ problem were used in \cite{KP-HT15} to show the optimal $\TwoExpTime$-upper bound for the satisfiability and finite satisfiability problems for the guarded fragment of \FOt+$\{E_1,E_2\}$.

The above approach has already been successfully applied to get optimal upper complexity bounds for extensions of 
\FOt{} where  the operation of {\em equivalence
	closure} can be applied to one or more binary predicates \cite{KMP-HT14,KP-HT15}. Such operators can be used to express non-first-order notions such as {\em reachability} or {\em connectedness} in undirected graphs---notions often encountered in practise. 

It remains open whether the linear/integer programming approach might be helpful in designing optimal decision procedures for logics with more than two variables, and in particular when the signatures feature predicates of higher arity.

\subsection{Remarks}
In this section we discussed several logics for which it required more care to proof decidability of the finite satisfiability problem than to prove decidability of the satisfiability problem. This by no means is a general trade. In particular, there are  logics such that \Sat{($\cal{L}$)} is undecidable and \FinSat{($\cal{L}$)} is decidable, or vice versa (see e.g.~\cite{MichOW12} for a family of examples from the elementary modal logics). 

We have also mentioned  fragments for which decidability of \FinSat{($\cal L$)} has been solved and the status of \Sat{($\cal L$)} remains open, this include some extensions of \FOt{} with order relations (cf.~\cite{SZ12,BDM11} for a detailed picture).

In this area one can also find fragments for which the complexity of the finite satisfiability problem jumps to classes like {\em vector addition systems} which are $\ExpSpace$-hard and are known to be decidable but no elementary upper bound has been found so far (cf.~\cite{Kosaraju82}). An example of this phenomenon is the extension of \Ct{} with one linear order and one successor of a linear order augmented with an additional binary relation studied in \cite{CW15}.

\section{Conclusion}
The picture concerning decidability of the (finite) satisfiability problems for extensions of fragments of first-order logic defined as the natural image of the standard translation of modal logic is multidimensional  and colourful. Current research in this area, apart from studying the open question already mentioned,  involves investigation of logics used by combining several operators from the already well understood fragments, identifying smaller fragments with better algorithmic properties, and optimization of known algorithms towards practical implementation. Finite model reasoning is crucial to both the theory and practice of computation. It is still not well understood when addressing the problem of query answering---the central reasoning problem of database theory. We believe that this problem will gain a lot of attention in the nearest future and will intensively use results from the areas outlined in this talk.



\bibliography{M4M}

\end{document}